# Simulating liquid-vapor phase separation under shear with lattice Boltzmann method

WANG Ce[1], XU Ai-Guo[2]†, ZHANG GuangCai[2] & LI YingJun[1]

[1] China University of Mining and Technology, Beijing 100083, China;
[2] National Key Laboratory of Compuataional Physics, Institute of Applied Physics and Computational Mathematics, Beijing 100088, China

**Abstract:** We study liquid-vapor phase separation under shear via the Shan-Chen lattice Boltzmann model. Besides the rheological characteristics, we analyze the Kelvin–Helmholtz (K-H) instability resulting from the tangential velocity difference of the fluids on two sides of the interface. We discuss also the growth behavior of droplets. The domains being close to the walls are lamellar-ordered, where the hydrodynamic effects dominate. The patterns in the bulk of the system are nearly isotropic, where the domain growth results mainly from the diffusion mechanism. Both the interfacial tension and the K-H instability make the liquid-bands near the walls tend to rupture. When the shear rate increases, the inequivalence of evaporation in the upstream and coagulation in the downstream of the flow as well as the role of surface tension makes the droplets elongate obliquely. Stronger convection makes easier the transferring of material particles so that droplets become larger.

Fluidic phenomena in nature are the statistical behaviors of the corresponding microsystems. The thermodynamic properties for a fluid system are determined by kinetic behaviors of the microsystems and the boundary conditions. It is impractical to track fully the evolutions of microsystems using the molecular dynamics (MD) simulation; at the same time, some macro-behaviors are not sensitive to some degrees of freedom. Traditional fluid dynamics does not work well for systems whose nonequilibrium effects are pronounced, for example, the multiphase system. Lattice Boltzmann (LB) method can represent partially the microscopic behavior of a realistic fluid system, and can also be associated with the macroscopic behavior. It is a mesoscopic simulation scheme. The LB method[1] is developed from the lattice vapor automata (LGA) method. Since the LB code has many advantages, for example, it is (i) simple, (ii) easy to deal with complex boundaries, (iii) easy for large-scale parallel computing, etc., it has been attracting more attention with time.

Phase separation of single-component and multicomponent fluids is a fundamental project. It is also a project having many applications in a variety of industrial and environmental processes, such as atmospheric quality improving, enhancing oil extraction, underground water purifications, etc.[2-10]. LBM is an important tool for simulating the single-phase and multiphase flow. Nowadays, the three well-known LB models for multiphase flows are the Rothman-Kellar model by Gunstensen, et al.[11] (1991), Pseudo-potential model by Shan and Chen[12] (1993), and free energy model by Swift et al.[13].

The multiphase LB model[14] by Shan and Chen can be used to simulate single-component multiphase flows and component flows. For the latter case, each compo-

†Corresponding author (email: Xu_Aiguo@iapcm.ac.cn)

Supported by the National Natural Science Foundation of China (Grant Nos. 10775018 and 10702010), the National Basic Research Program of China (Grant No. 2007CB815105), and the Science Foundations of LCP and CAEP nent can be immiscible and have different masses. The equilibrium state of each component has a nonideal-vapor equation of state. This model introduces interaction forces between particles. Due to collision rules in this model are local, it is highly efficient to compute on massively parallel computers. This model suits for large-scale numerical simulations of various types of fluid flows. Study on mulitphase flows and phase separation has a long history[15]. But liquid-vapor phase separation has not got so extensive studies as the phase separations of binary fluid systems. In this study, we apply the Shan-Chen model to study the liquid-vapor phase separation. Besides the rheological behaviors, we analyze the Kelvin-Helmholtz (K-H) instability resulting from the tangential velocity difference of the fluids on the two sides of the interface. We discuss also the growth behavior of droplets.

## 1 The Shan and Chen model

We use the Shan-Chen LB model based on the D2Q9 lattice. The velocity vectors are as follows:

$$
\begin{aligned}
&\mathbf{e}_i = (0,0), i=0 \\
&\mathbf{e}_i = \left(\cos\frac{(i-1)\pi}{2}, \sin\frac{(i-1)\pi}{2}\right)c, i=1,...4 \\
&\mathbf{e}_i = \left(\cos\frac{(2i-1)\pi}{4}, \sin\frac{(2i-1)\pi}{4}\right)\sqrt{2}c, i=5,...8
\end{aligned} \quad (1)
$$

where $c$ is the propagation velocity. The particle's distribution function is $f_i(\mathbf{x},t)$, and its local equilibrium distribution function is $f_i^{eq}(\mathbf{x},t)$. Under the condition without external force, evolution of the system follows lattice Boltzmann equation:

$$f_i(\mathbf{x}+t,t+1) - f_i(\mathbf{x},t) = -\frac{1}{\tau}[f_i(\mathbf{x},t) - f_i^{eq}(\mathbf{x},t)], \quad (2)$$

where $\tau$ is a single relaxation-time parameter, and the local equilibrium distribution functions are

$$
\begin{aligned}
f_0^{eq} &= \frac{4}{9}\rho\left[1 - \frac{(\mathbf{u}\cdot\mathbf{u})^2}{2T}\right], \\
f_i^{eq} &= \frac{1}{9}\rho\left[1 + \frac{(\mathbf{e}_i\cdot\mathbf{u})}{T} + \frac{(\mathbf{e}_i\cdot\mathbf{u})^2}{2T^2} - \frac{(\mathbf{u}\cdot\mathbf{u})}{2T}\right], i=1,...4, \quad (3)\\
f_i^{eq} &= \frac{1}{36}\rho\left[1 + \frac{(\mathbf{e}_i\cdot\mathbf{u})}{T} + \frac{(\mathbf{e}_i\cdot\mathbf{u})^2}{2T^2} - \frac{(\mathbf{u}\cdot\mathbf{u})}{2T}\right], i=5,...8,
\end{aligned}
$$

where $T$ is the temperature of the system, here $T=c^2/3$.

In order to describe the non-ideal vapor or multiphase system, we consider the interaction potential between particles, $V = -G\psi^2/2$, where $\psi = \rho_0(1-\exp(-\rho/\rho_0))$ is an effective density, $\rho_0$ is the reference density, $G$ is the interaction parameters. Therefore, the interacting force $F = G\psi\nabla\psi$, where the sign of $G$ determines the force is attractive or repulsive. We add external force to the collision term[16]:

$$
\begin{aligned}
&f_i(\mathbf{x}+\mathbf{c}_it,t+1) - f_i(\mathbf{x},t) \\
&= -\frac{1}{\tau}[f_i(\mathbf{x},t) - f_i^{eq}(\mathbf{x},t)] + F_i.
\end{aligned} \quad (4)
$$

Due to that the velocity vector is fixed in the LB equation, there is a constraint to the external force term in order to ensure the conservations of density and momentum.

$$F_i = w_i \mathbf{e}_i \cdot \mathbf{F}/T, \quad (5)$$

where $w_i$ is the weight factor in different directions.

By applying the Chapman-Enskog expansion[17], we can obtain the following Navier-Stokes equations:

$$
\begin{aligned}
&\frac{\partial\rho}{\partial t} + \nabla\cdot(\rho\mathbf{u}) = 0, \\
&\frac{\partial(\rho\mathbf{u})}{\partial t} + \nabla\cdot(\rho\mathbf{u}\mathbf{u}) = -\nabla p + \nu\nabla\cdot[\rho((\nabla\mathbf{u})+(\nabla\mathbf{u})^T)],
\end{aligned} \quad (6)
$$

where $\nu = (\tau-1/2)c^2\Delta t/3$ is the kinematic viscosity, the equation of state reads:

$$p = \rho T - \frac{1}{2}G\psi^2(\rho). \quad (7)$$

Obviously, the pressure is not a monotonic increasing function of density, thermodynamic phase transition may occur in such a system. By setting the first and second orders derivatives of $p$ to be zero:

$$
\begin{aligned}
&\frac{\partial p}{\partial \rho} = 0, \\
&\frac{\partial^2 p}{\partial \rho^2} = 0,
\end{aligned} \quad (8)
$$

we can obtain the critical point $G_c$. From the above two equations, we get the following relations: When $\rho_C = \rho_0 \ln 2$, $T_C = \frac{G\rho_0}{4}$ or $G_C = \frac{4T}{\rho_0}$. It is clear that when the temperature is lower than $T_C$ or the interaction parameter $G > G_C$, liquid-vapor phase separation occurs.

## 2 The shear boundary condition

In order to enforce a shear flow on the system, we have used the following scheme[3]. We assume that the shear follow is directed along the horizontal direction. The upper wall velocity is $u = \gamma D/2$ and move to the right, where $D$ is the distance between the two walls. The lower wall moves to the left. Let us consider the upper wall (similar considerations apply to the lower wall). After the propagation the distribution functions $f_1(x,t)$, $f_2(x,t)$, $f_6(x,t)$, $f_5(x,t)$ and $f_3(x,t)$ are known on each site, while $f_0(x,t)$, $f_4(x,t)$, $f_7(x,t)$, $f_8(x,t)$ are not.

From $\sum_i f_i = n$, we get

$$f_4(x,t) + f_7(x,t) + f_8(x,t) = n - (f_0(x,t) + f_1(x,t) + f_2(x,t) + f_3(x,t) + f_5(x,t) + f_6(x,t)). \quad (9)$$

From $\sum_i f_i \mathbf{e}_i = n\mathbf{u}$, we get

$$f_8(x,t) - f_7(x,t) = n\gamma D/2c \\ -[f_1(x,t) + f_5(x,t) - f_6(x,t) - f_3(x,t)] \quad (10)$$

in the $x$ direction and

$$f_4(x,t) + f_7(x,t) + f_8(x,t) \\ = f_2(x,t) + f_5(x,t) + f_6(x,t) \quad (11)$$

in the $y$ direction, where $\gamma$ is the shear rate. The equation system has three equations with four unknown variables. To close the system of equations the bounceback rule[3] is adopted for the distribution functions normal to the boundary. That is

$$f_2(x,t) = f_4(x,t). \quad (12)$$

In order to preserve correctly mass conservation, we add a further constraint. That is,

$n = \hat{n}$
$$\hat{n}(x, t-\Delta t) = f_0(x, t-\Delta t) + f_5(x, t-\Delta t) + f_2(x, t-\Delta t) \\ + f_6(x, t-\Delta t) + f_1(x,t) + f_3(x,t) \\ + f_2(x,t) + f_5(x,t) + f_6(x,t), \quad (13)$$

where quantities at time $(t-\Delta t)$ refer to the previous time step and have not been propagated over the lattice. In order to constrain that on all boundary sites $n = \hat{n}$, we have to introduce an independent variable in the system of equations. We choose $f_0(t)$ since it does not propagate. Therefore,

$$f_0(x,t) = \hat{n} - [f_1(x,t) + f_3(x,t)] \\ - 2[f_2(x,t) + f_5(x,t) + f_6(x,t)],$$
$$f_4(x,t) = f_2(x,t), \quad (14)$$
$$f_7(x,t) = f_5(x,t) + \frac{1}{2}[f_1(x,t) - f_3(x,t)] - n\gamma D/2c,$$
$$f_8(x,t) = f_6(x,t) + \frac{1}{2}[f_3(x,t) - f_1(x,t)] + n\gamma D/2c.$$

## 3 Simulation results

With the LB model in sec.1, we choose $T = 1/3$, $\rho_0 = 1/\ln 2$, while $G_C \approx 0.92$. The initial density of system is around 1 with small fluctuations. The lattice size is 256×256, period boundary conditions are used in the horizontal direction and shear boundary conditions are used in the vertical direction. In simulations, we choose $G \approx 0.95 > G_C$ (the corresponding critical temperature is $T_C \approx 0.34$, it is clear that $T = 1/3 < T_C$).

### 3.1 Rheological characteristics of liquid-vapor phase

This section is focused on the shear effects on the rheological characteristics of the whole system. What we consider is a temperature quench, which means that the dynamic details occurring in the temperature-decreasing procedure is not taken into account. In Figures 1 and 2 the slight color is for liquid phase, and the deep color is for vapor. The initial state of the system is set as a constant with tiny fluctuations, corresponding to high temperature uniform single-phase. When the temperature abruptly falls below the critical point and the system enters the parameter regime for two-phase coexistence, the density fluctuations in the system cause the droplets to grow up. Each droplet has a tendency to absorb molecules form the surroundings, in other words, it has the tendency to grow up by diffusion. Since fewer and larger droplets can decrease the interfacial energy, with the evolution, small droplets disappear gradually and the bigger droplets grow with time. The means size of domains of the system increase gradually. As the boundary flow, the droplets nearby can integrate easily with liquid-bands. Since shear effects have not come into the bulk of the system, the domains there are still nearly isotropic. With time going on, the liquid-bands near the walls absorb molecules from the inside and begin to become wider. Both the interfacial tension and the K-H instability make the liquid-bands near the walls tend to rupture. See Figure 1.

With the increase of shear rate, the anisotropic characteristics is more pronounced. Figure 2 shows density configurations for the case of $\gamma = 10^{-4}$. Compared with Figure 1, we can see two differences: when the shear rate increases, the stronger convection makes easier the transferring of material particles so that droplets become larger; the unbalance of evaporation in the upstream and condensation in the downstream of the flow makes the droplets elongated obliquely (discuss in details in sec. 3.4 and 3.5).

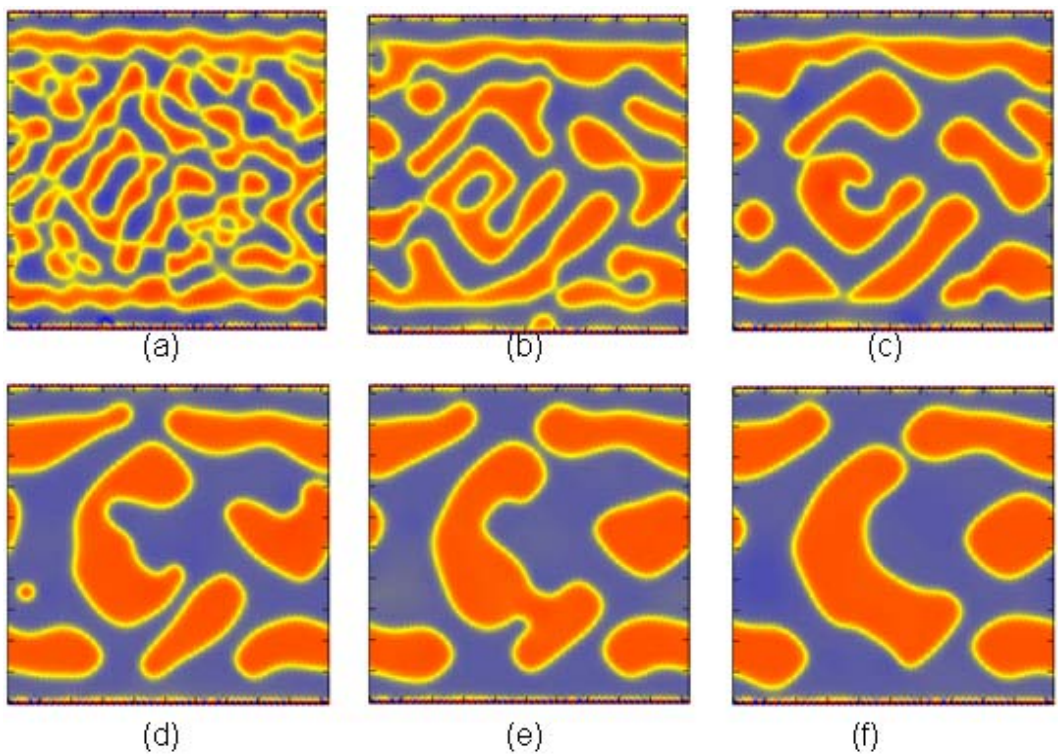

**Figure 1** Snapshots for phase separation, where $\gamma = 10^{-5}$. (a)—(f) $t = 500, 1000, 1500, 2000, 2500$ and $3000$, respectively.

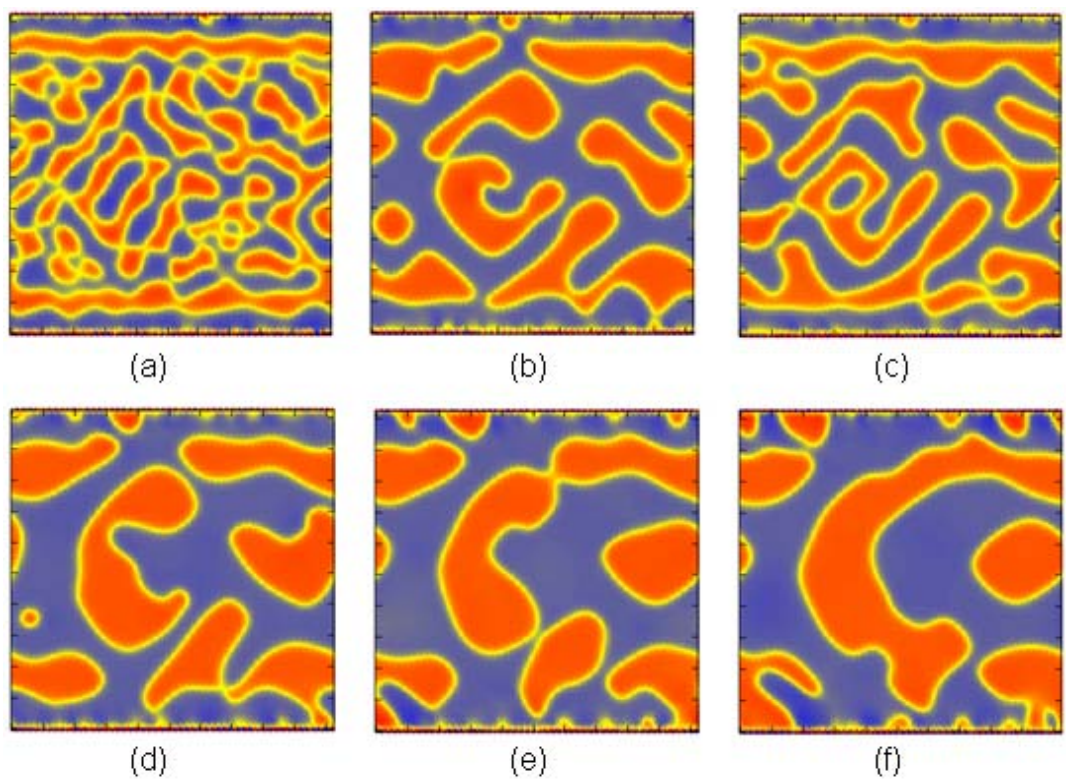

**Figure 2** Density configurations for $\gamma = 10^{-4}$. (a)—(f) $t = 500, 1000, 1500, 2000, 2500$ and $3000$, respectively.

The dynamical characteristics of the droplets can be seen via their structure factors. Figure 3 shows the structure factors corresponding to the configurations in Figure 2. In Figures 1 and 2, domains being close to the walls are lamellar-ordered since the hydrodynamic effects dominate there; the patterns in the bulk of the system are nearly isotropic since the domain growth results mainly from the diffusion mechanism. In Figure 3, we can find that, when $t<500$, the shear makes only some effects close to the walls, correspondingly, the structure factor is nearly circular and isotropic. With time going on, the structure factor becomes elliptic. When $t=2000$ two peaks appear at $k_x=0$ along the flow direction. Then the shear effects come into the system. After that the elliptic structure becomes thinner, and the two peaks become more pronounced.

### 3.2 Kelvin–Helmholtz (K-H) instability

In Figures 1 and 2 vapor-bands (the deep color is for vapor) partly prevent the shear effect from entering the system. Then the main transport mechanisms are diffusion and local flow. Near the boundary walls, the flow bands on the two sides of the interfaces have tangential velocity differences, see Figure 4. The tangential velocity difference is the reason for the K-H instability. Together with the interfacial tension, the K-H instability makes thicker the thicker parts and makes thinner the thinner parts of the liquid bands. Finally the liquid bands rupture.

### 3.3 The mechanism for spherical droplets

In order to study the dynamic characteristics of the droplets, we calculate the spatial distribution of changing rate of density, denoted by $v$, in Figure 5(a). From Figure 5(a) it is clear that when the droplets' surface curvature is negative, corresponding changing rate of density is positive. The density increasing shows that the vapor in the vicinity of the droplets has a coagulation trend; when the droplets' surface curvature is positive, corresponding changing rate of density is negative. The density decrease shows that evaporation occurs at the surface of droplets. Surface tension makes the droplets' surface flatter and flatter. In order to understand better the relationship between congealment (or vaporization) and droplets' surface curvature, we show partial enlarged plots with particle velocity fields in Figures 5(b) and 5(c). Figures 5(b) and 5(c) correspond to regimes in square 1 and square 2 in Figure 5(a). The droplets' surface curvatures are positive and negative, respectively. From Figure 5(b) it is clear that material particles gather to a place, where the droplets' surface curvature is negative. From Figure 5(c) we find a reverse process where material particles diverge form a place where the droplets' surface curvature is positive. In Figure 1 we see a round droplet when $t=2000$, its coordinate is from 1 to 25 in the $x$ direction, and from 80 to 100 in the $y$ direction. Its surface curvature is large, and corresponding changing rate of density is negative, therefore evaporation occurs there. The droplet has completely evaporated when $t=2500$.

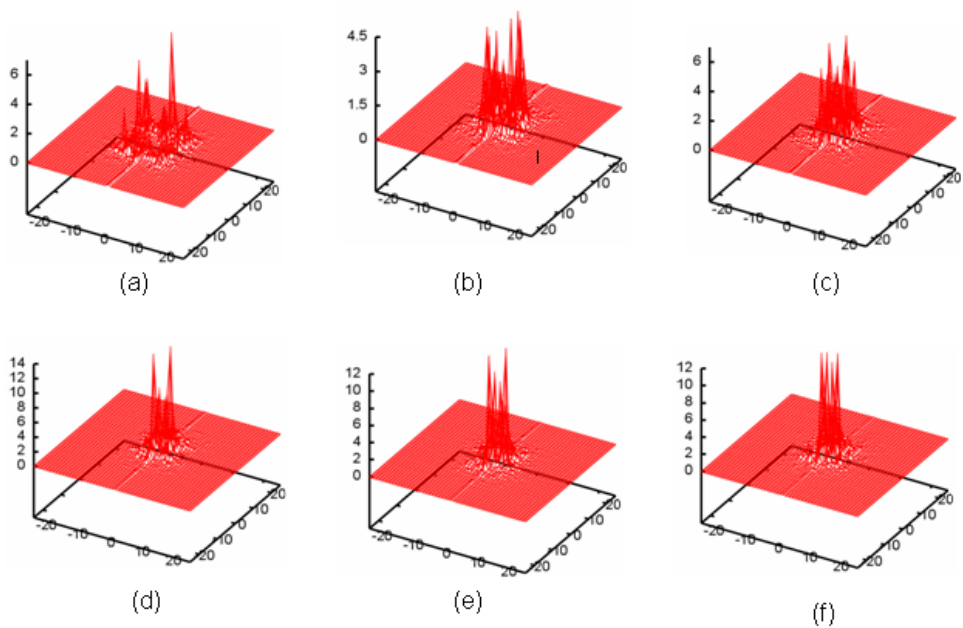

**Figure 3** Structure factors for $\gamma=10^{-4}$. (a)−(f) $t=500, 1000, 1500, 2000, 2500$ and $3000$, respectively. The units of $k_x$ and $k_y$ are $2\pi/256$.

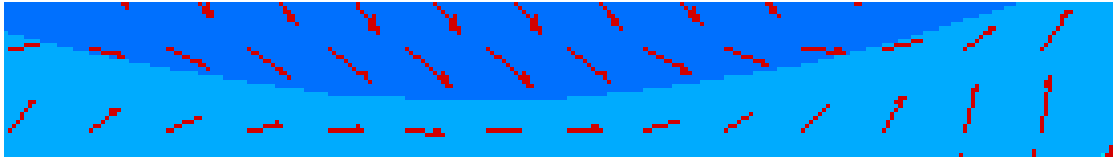

**Figure 4** The Configuration with the velocity field near a interface, where $t=500$.

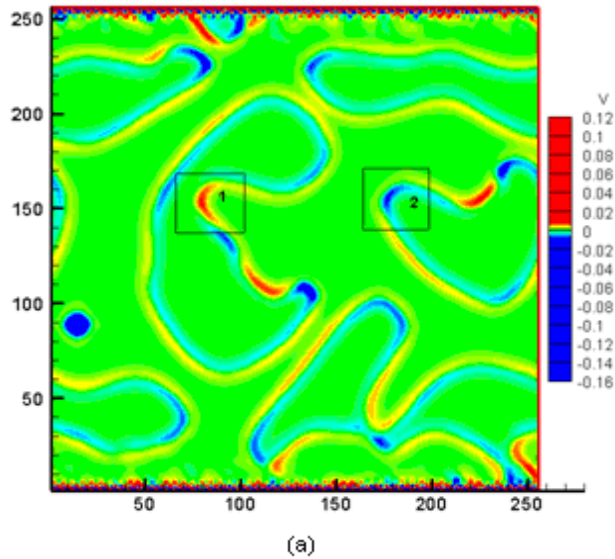

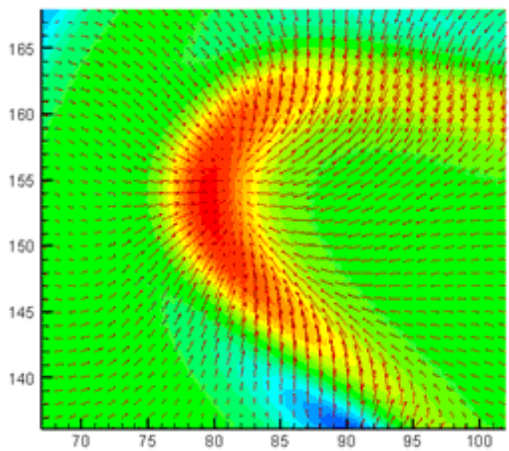

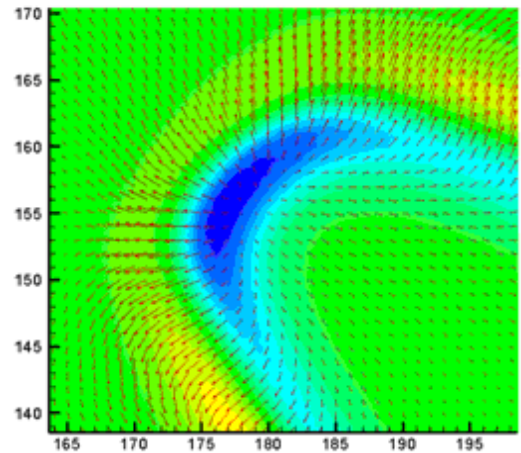

**Figure 5** The changing rate of density with the velocity field, where $t=2000$. (a) the changing rate of density; (b) the portion in square 1 of (a); (c) the portion in square 2 of (a). The unit here is lattice unit.

### 3.4 The mechanism of band forming

In Figures 6(a)–(c) the slight color is for liquid, and the deep color is for vapor. When the shear rate increases, the banded structure becomes more pronounced with time, see Figure 6(a). After some time the liquid forms an obliquely banded structure in Figure 6(c). The oblique elongation of band structure results from the unbalance of evaporation in the upstream and coagulation in the downstream of the flow. Figures 6(d)–(f) show the changing rates of density corresponding to Figures 6(a)–(c). The changing rate of density of the liquid is negative in the upstream and is positive in the downstream.

### 3.5 The mechanism of droplets on the walls

Compared with Figures 1 and 2, we can see that, when the shear rate increases, it becomes easier for the droplets to grow on the walls. Although droplets on walls can only obtain molecules from one side, stronger convection makes material particles collide and integrate more frequently. Thus, it is easier for the droplets to grow. By tracking the evolution of domains, we can find that, the droplet on the upper wall in Figure 2, located within $150 \leqslant x \leqslant 175$ at time $t=500$, can absorb material particles and grow up more easily when convection becomes stronger. The liquid bands near the walls are controlled simultaneously by the surface tension and the K-H instability. Because the K-H instability dominates, the liquid-bands rupture finally.

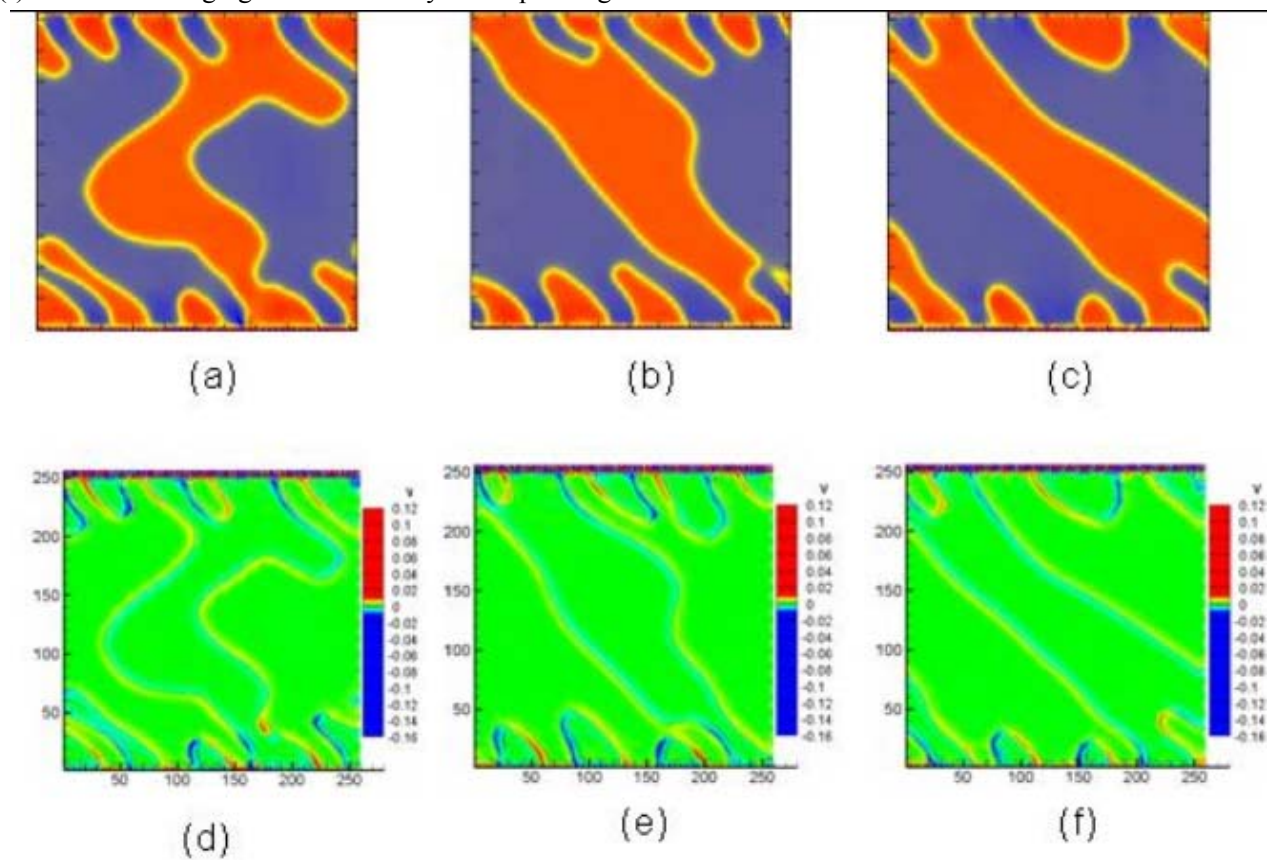

**Figure 6** Density and changing rates of density configurations. (a)–(c) The configurations for $\gamma=10^{-4}$ at times $t=7000, 16000, 25000$; (d)–(f) the changing rates of density corresponding to (a)–(c).

## 4 Conclusion

We apply the Shan-Chen lattice Boltzmann model to study the liquid-vapor phase separation under shear. We focus on the rheological characteristics, the K-H instability resulting from the tangential velocity difference of the fluids on the two sides of the interface and the growth behavior of droplets. Because the vapor phase transfers the shear effect very weekly, the patterns in the bulk of the system are nearly isotropic, where the domain growth results mainly from the diffusion mechanism. Both the interfacial tension and the K-H instability make the liquid-bands near the walls tend to rupture. When the shear rate increases, the unbalance of evaporation in the upstream and condensation in the downstream of the flow, together with the surface tension, makes the domains elongate obliquely. Stronger convection makes easier transferring of material particles so that droplets become easier to be larger. In the future work, we will investigate the effects of viscosity, oscillatory shear, etc. on the rheological behaviors liquid-vapor phase separation. We hope this work will inspire more experimental work.

*The authors would like to thank Gan Yanbiao, Wang Lifeng, and Chen Feng for helpful discussion.*